\documentclass[10pt,a4paper,conference]{IEEEtran_souby}
\IEEEoverridecommandlockouts
\renewcommand{\baselinestretch}{1}

\usepackage{cite}
\usepackage{enumitem}
\usepackage{multirow}
\usepackage{wrapfig}
\usepackage{mdwlist}
\usepackage{xspace}
\usepackage{url}
\usepackage{epstopdf}
\usepackage{graphicx}
\usepackage{caption}
\usepackage{xcolor}
\usepackage{anyfontsize}
\usepackage[ruled,vlined]{algorithm2e}
\usepackage{etoolbox}
\usepackage{times}
\usepackage{amsmath}
\usepackage{textcomp}
 
\newcommand{\etc}{\textit{etc.,}\xspace}
\newcommand{\eg}{\textit{e.g.,}\xspace}
\newcommand{\ie}{\textit{i.e.,}\xspace}
\newcommand{\combL}{combination PUF\xspace}

\newcommand{\comb}{\emph{C-PUF}\xspace}

\newcommand{\ins}{instance\xspace}
\newcommand{\inss}{instances\xspace}
\newcommand{\cins}{\emph{C-PUF} instance\xspace}
\newcommand{\cinss}{\emph{C-PUF} instances\xspace}
\newcommand{\red}{\color{black}}

\graphicspath{{./FiguresInPaper/}}

\begin{document}
	

\title{Memory-based Combination PUFs for Device Authentication in Embedded Systems
}

\author{
		\IEEEauthorblockN{Soubhagya Sutar, Arnab Raha, and Vijay Raghunathan}
        \IEEEauthorblockA{School of Electrical and Computer Engineering, Purdue University}
        \IEEEauthorblockA{\{ssutar,araha,vr\}@purdue.edu}
}


\maketitle

\begin{abstract}
Embedded systems play a crucial role in fueling the growth of the Internet-of-Things (IoT) in application domains such as health care, home automation, transportation, \emph{etc}. However, their increasingly network-connected nature, coupled with their ability to access potentially sensitive/confidential information, has given rise to many security and privacy concerns. An additional challenge is the growing number of counterfeit components in these devices, resulting in serious reliability and financial implications. Physically Unclonable Functions (PUFs) are a promising security primitive to help address these concerns. Memory-based PUFs are particularly attractive as they require minimal or no additional hardware for their operation. However, current memory-based PUFs utilize only a single memory technology for constructing the PUF, which has several disadvantages including making them vulnerable to security attacks. In this paper, we propose the design of a new memory-based combination PUF that intelligently combines two memory technologies, SRAM and DRAM, to overcome these shortcomings. The proposed combination PUF exhibits high entropy, supports a large number of challenge-response pairs, and is intrinsically reconfigurable. We have implemented the proposed combination PUF using a Terasic TR4-230 FPGA board and several off-the-shelf SRAMs and DRAMs. Experimental results demonstrate substantial improvements over current memory-based PUFs including the ability to resist various attacks. Extensive authentication tests across a wide temperature range (20$^{\circ}$C- 60$^{\circ}$C) and accelerated aging (12 months) demonstrate the robustness of the proposed design, which achieves a 100\% true-positive rate and 0\% false-positive rate for authentication across these parameter ranges.
\end{abstract}

 \section{Introduction}
\label{sec:intro}
The Internet-of-Things (IoT) is one of the fastest growing technologies across all of computing, revolutionizing a number of application domains such as industrial manufacturing, home automation, wearable computing, \emph{etc}. 
However, this rapid proliferation has brought with it a plethora of new security and privacy concerns~\cite{invasive,ml_attack}. 
Further, with hardware components (ICs and IP cores) being sourced from manufacturers across the globe, instances of counterfeiting/piracy have increased steadily, leading to serious reliability implications and substantial revenue loss (over \$100 billion annually \cite{bogus}).

\begin{figure}[t]
	\centering
	\includegraphics[width=0.9\columnwidth]{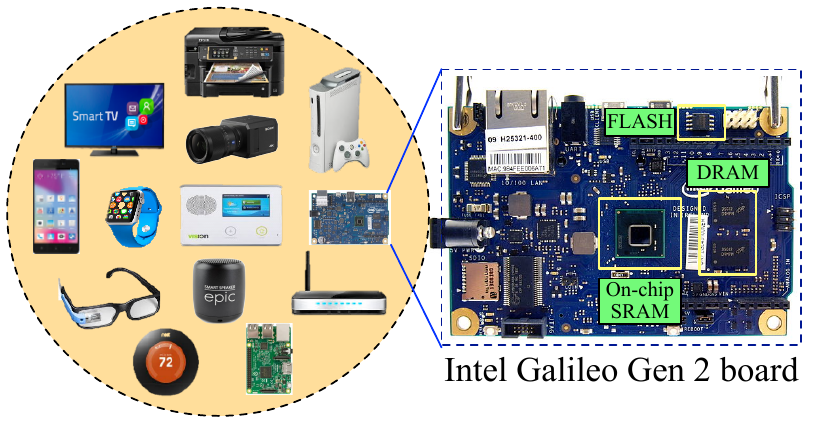}
	\vskip -4pt
	\caption{IoT devices featuring multiple memory technologies}
	\label{fig:iot}
	\vskip -15pt
\end{figure}

Hardware-intrinsic security mechanisms such as Physically Unclonable Functions (PUFs) offer a secure, low-cost, and robust solution for addressing these challenges \cite{dev_auth}. PUFs exploit the random variations inherent in the manufacturing process to extract unclonable and instance-specific keys (fingerprints) from hardware components.
The instance-specific nature of the keys enables us to uniquely identify and authenticate each device \cite{dev_auth, dpuf_cases2} based on a challenge-response mechanism, thereby addressing the problems of access control as well as counterfeiting. While a variety of PUFs have been proposed, memory-based PUFs \cite{sram_puf_holcomb, dram_puf_keller, dpuf_cases2, commodity_dpuf, flash_puf_prabhu}, in particular, are  attractive options due to the ubiquitous presence of memory in every embedded device (Fig.~\ref{fig:iot}). Further, they require minimal (or no) additional hardware, unlike other PUF implementations. 

Current memory-based PUFs, however, suffer from several shortcomings such as low entropy~\cite{dram_puf_keller, flash_puf_prabhu}, limited number of Challenge-Response Pairs (CRPs)~\cite{sram_puf_holcomb, flash_puf_prabhu}, susceptibility to environmental and temporal variations (requiring complex error correction)~\cite{dram_puf_keller, dpuf_cases2}, and high operational latency~\cite{dram_puf_keller, flash_puf_prabhu}. Most importantly, current memory-based PUFs are constructed using a single memory component (or technology) in the device, \ie based on a single entropy source. If the memory component is removable from the system ({\em e.g.,} a DRAM SODIMM) and is transferred to a different system, the identity transfers over as well, which is undesirable. To mitigate this, it is desirable that the PUF be dependent on multiple system components (some of which may be more tightly integrated, and thus harder to remove, than others).

Recent works \cite{dist_puf, super_puf} have addressed a subset of these shortcomings. However, they also require the addition of custom hardware to the system and, hence, cannot be implemented using Commercial-Off-The-Shelf (COTS) systems. In this paper, we overcome these limitations by proposing the design of a memory-based \combL (henceforth referred to as \comb). \comb intelligently utilizes two widely-used memory technologies, Static Random Access Memory (SRAM) and Dynamic Random Access Memory (DRAM), to construct a PUF, thereby synergistically combining the advantages of both types of memory PUFs, while addressing the various shortcomings associated with single memory based PUFs mentioned above. The heterogeneous nature of the entropy sources (memories) used and \comb's ability to undergo \emph{intrinsic reconfiguration} (ability to reconfigure the PUF at runtime without any additional hardware) protects it from various security attacks. \comb also features two lightweight error-correction algorithms to ensure robust operation (authentication) even under wide environmental and temporal variations. Specifically, this paper makes the following contributions:

\begin{itemize}
	
\item We propose the concept and design of a memory-based combination PUF (\comb) that intelligently uses two memory technologies, SRAM and DRAM, to construct a PUF that~\emph{(i)} exhibits high entropy and supports a large number of CRPs, \emph{(ii)} is intrinsically reconfigurable, \emph{(iii)} is robust to environmental and temporal variations, and \emph{(iv)} does not require additional (custom) hardware, hence, can be natively implemented on a COTS device.

\item As a key enabler for \comb, we propose two lightweight algorithms for performing error correction in SRAM start-up values. These algorithms enable us to achieve perfect error correction of SRAM bit-errors, thereby ensuring robust operation (authentication) under environmental and temporal variations.

\item We implement, demonstrate, and evaluate a fully-functional prototype of \comb in a real system using several off-the-shelf SRAMs and DRAMs. Extensive authentication tests performed across a wide temperature range (20$^{\circ}$C - 60$^{\circ}$C) and accelerated aging (12 months) achieved a 100\% true-positive rate and 0\% false-positive rate for authentication, demonstrating the robustness of the proposed design.
\end{itemize}
 \vspace{-0.05in}
\section{Background and Motivation}
\label{sec:bg}
Next, we provide a brief background on challenge-response-based authentication and memory-based PUFs, followed by motivating the need for the proposed combination PUF.


\vspace{-0.05in}
\subsection{{\red Challenge-response-based Authentication using a PUF}}
\label{bg:auth}
Device authentication is the process by which a trusted system (authenticator) verifies the identity of an untrusted device (client) before granting it access to any data or resources. It is usually performed using a challenge-response mechanism \cite{dev_auth, dpuf_cases2}. To verify the identity of a client, the authenticator first provides it with a challenge. The client then generates a response to the challenge using its on-board PUF. Prior to this, the authenticator creates a Challenge-Response Pair (CRP) database that stores all the challenges and their expected responses from genuine clients. By comparing the current client's response against the one stored in the CRP database, the authenticator infers whether the client is genuine or not. 

\vspace{-0.05in}
\subsection{{\red SRAM and DRAM PUFs}}
\label{bg_sram}
\begin{figure}
	\centering
 	\includegraphics[width=1\columnwidth]{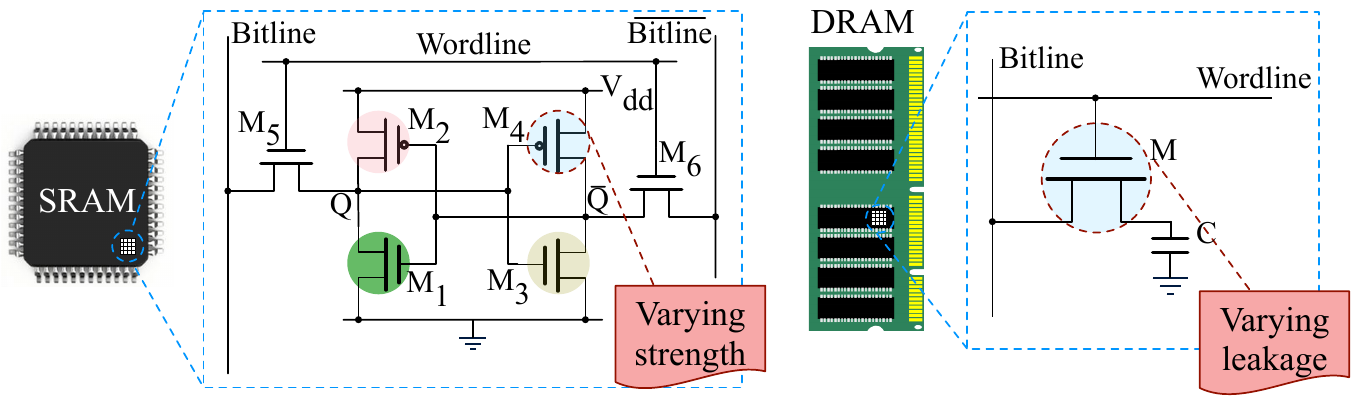}
    \vskip -2pt
	\caption{SRAM and DRAM bit cells}
	\label{fig:sram-dram}
	\vskip -13pt
\end{figure}

Each cell or bit in an SRAM is arranged in a six-transistor configuration (most common)
consisting of cross-coupled CMOS inverters (M\textsubscript{1}\textendash M\textsubscript{4}) and access transistors (M\textsubscript{5}\textendash M\textsubscript{6}), as shown in Fig.~\ref{fig:sram-dram} (left). Powering-up the SRAM causes each cell to reach one of two states (start-up values), [$Q$=1, $\overline{Q}$=0] or [$Q$=0, $\overline{Q}$=1], depending upon the relative strengths of the transistors as well as noise. 
Process variations during manufacturing cause these strengths to vary across SRAMs, leading to different start-up values for different SRAMs. This forms the foundation of an SRAM PUF \cite{sram_puf_holcomb, mc_spuf} that uses the \emph{power-cycling} (power off $\rightarrow$ power on $\rightarrow$ read SRAM) approach to generate unique start-up values as responses.

Fig.~\ref{fig:sram-dram} (right) shows the fundamental building blocks of a DRAM bit cell, namely an access transistor (M) and capacitor (C). The bit-value is decided by the charge on the capacitor; full charge implies `1' and no charge implies `0', or vice-versa. This charge leaks over time,
eventually resulting in the loss of data stored in the cell, which is referred to as a \emph{bit-flip} (`1'$\rightarrow$`0' or `0'$\rightarrow$`1'). To prevent this, the DRAM memory-controller refreshes the cells (replenishes the charge) periodically (\eg every 64 ms). Due to process variations, the rate of leakage (or bit-flip) varies widely across DRAMs (and within the same DRAM). This forms the basis of the \emph{refresh-pausing} approach in a DRAM PUF~\cite{dram_puf_keller, dpuf_cases2}, in which refresh operations are (intentionally) paused for a certain time-interval, generating unique bit-flip patterns in the DRAM data. This data is then read out and forms the PUF's response. 
\subsection {Motivation}
\label{sec:mot}
Each of the PUFs described above has shortcomings. An SRAM PUF exhibits high entropy but supports a small number of CRPs due to the existence of very few variable parameters (in its challenge-response mechanism) as well the small extent to which these parameters could be varied because of an SRAM's usually small size (capacity) in a system.
On the other hand, a DRAM PUF employs a challenge-response mechanism involving several widely-variable parameters, supporting a large number of CRPs. However, for practical \emph{refresh-pause intervals}, the entropy (and uniqueness) exhibited by it is much lower than an SRAM PUF. 
Also, DRAM is often loosely integrated in a system (\eg using a removable/replaceable DRAM SODIMM). If the DRAM SODIMM is removed and transferred to a different system, the identity transfers over as well, which is undesirable. These shortcomings motivate the design of the proposed \comb.

 \begin{figure*}[t]
	\centering
	\includegraphics[width=1\hsize]{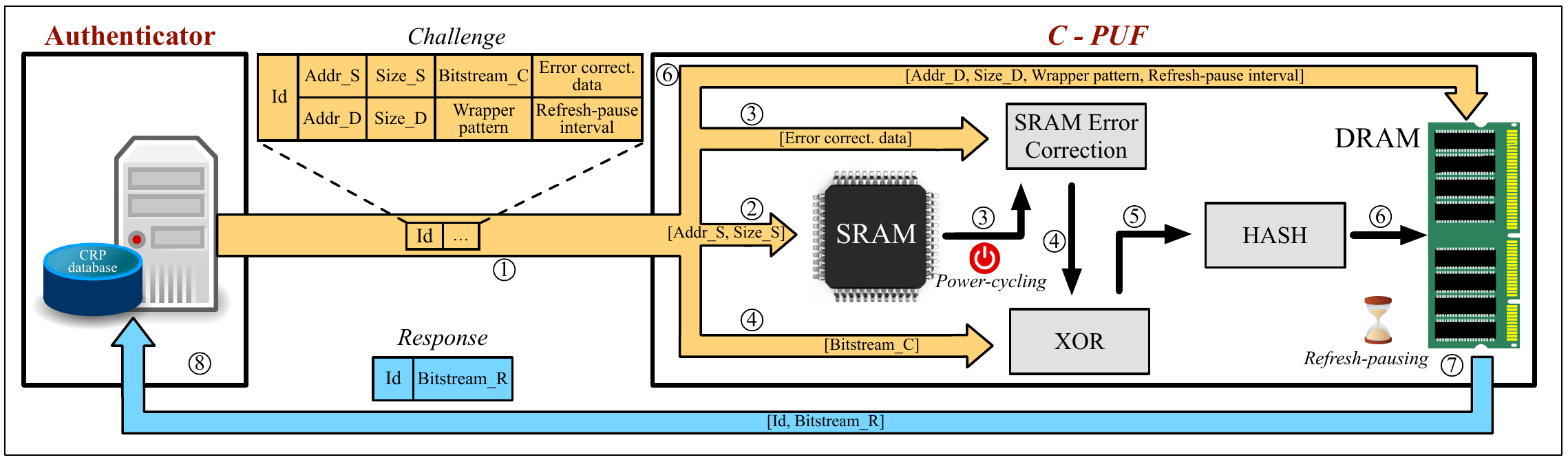}
	\caption{Overview of the proposed \comb architecture}
	\label{fig:arch}
	\vskip -15pt
\end{figure*}

\section{C-PUF Architecture and Design}
\label{sec:des}
\comb is designed to perform challenge-response-based authentication in a device.
As shown in Fig.~\ref{fig:arch}, it utilizes two widely-used memory technologies, SRAM and DRAM, that act as heterogeneous sources of entropy.
Note that the SRAM (PUF) utilizes the \emph{power-cycling} approach to generate start-up values as responses while the DRAM's (PUF) responses comprise of unique bit-flip patterns generated through the \emph{refresh-pausing} approach. In \comb, we tightly couple these two approaches using two mathematical operations (stages), \emph{XOR} and \emph{HASH} (explained later), to form \emph{C-PUF's} final response. {\red This combination allows the (challenge-response) behavior of one entropy source to influence that of the other in an unpredictable manner; the result is a PUF with high entropy and an exponential number of CRPs, which is much higher than what is supported by a standalone SRAM and DRAM PUF together (Sec.~\ref{sec:dis}).
Note that SRAM is usually tightly integrated with the processor and located on-the same chip/die, whereas DRAM is usually more loosely integrated (\eg as an external SODIMM). By combining these two components in \comb, we authenticate both an on-chip and off-chip component, thereby taking a step towards multi-component authentication in a device. Most importantly, all this is achieved without incorporating any additional (custom) hardware, and hence the design can be easily implemented on a COTS device.} Next, we present the architecture and design of \comb in detail. 


\subsection{Challenge-Response Mechanism}
\label{sec:cr_mech}
\comb employs a challenge-response mechanism that utilizes both SRAM \emph{power-cycling} and DRAM \emph{refresh-pausing}. The formats of the challenge and response used by \comb are depicted in Fig.~\ref{fig:arch}; while some parameters in the challenge are SRAM-specific, others are DRAM-specific except \emph{Id}, which represents a unique identifier assigned to a challenge and its corresponding response. The numbers specified against the arrows in Fig.~\ref{fig:arch} specify the sequence of operations during the associated response-generation process. To begin with, the authenticator sends a challenge to \comb in the proper format [step \textcircled{\raisebox{-.9pt}1}].
The SRAM then undergoes \emph{power-cycling}, as described in Sec.~\ref{bg_sram}, to generate a start-up value of \emph{Size\textunderscore S} bits from a block beginning at address \emph{Addr\textunderscore S} [step \textcircled{\raisebox{-.9pt}2}]. This start-up value is then corrected for bit-errors in the \emph{SRAM Error Correction} stage with respect to a previously generated \emph{golden} (or expected) start-up value using Algo.~\ref{algo:two} (Sec.~\ref{sec:err}) [step \textcircled{\raisebox{-.9pt}3}]. The information required for this correction is contained in the \emph{Error correction data} field of the challenge and is generated (prior to this) using Algo.~\ref{algo:one} (Sec.~\ref{sec:err}). Note that it is the same \emph{SRAM Error Correction} stage that is responsible for generating \emph{Error correction data} as well as performing actual error correction; this shall become clearer in the following paragraphs.
Next, the \emph{XOR} stage repeatedly applies the (bit-wise) mathematical operation~\textendash~\emph{xor} to the corrected start-up value (\emph{CV}), generated in the previous stage, and \emph{Bitstream\textunderscore C}; \emph{Bitstream\textunderscore C} is a random binary sequence of \emph{Size\textunderscore D} bytes  and is \emph{xor}-ed across its entire length with \emph{CV} [step \textcircled{\raisebox{-.9pt}4}]. The \emph{xor}-ed value then moves to the \emph{HASH} stage, where it is broken down into equally-sized chunks (\eg 32 bytes), each of which undergoes a mathematical \emph{hash} operation using SHA-256 [step \textcircled{\raisebox{-.9pt}5}]. The output from each chunk is concatenated together to form the complete \emph{hash}-ed value (\emph{HV}). Hashing helps to mask the SRAM start-up value and adds another layer of protection against attacks. 

Next, \emph{HV} is applied to the DRAM alongside other parameters \emph{viz.} \emph{Addr\textunderscore D}, \emph{Size\textunderscore D}, \emph{Wrapper pattern}, and \emph{Refresh-pause interval}, to undergo \emph{refresh-pausing}; we follow a similar methodology as described in \cite{dpuf_cases2} in this stage [step \textcircled{\raisebox{-.9pt}6}]. Specifically, the \emph{HV} (of \emph{Size\textunderscore D} bits) along with the peripheral data, specified by \emph{Wrapper pattern} (explained below), is first written onto a block in the DRAM, whose location is specified by \emph{Addr\textunderscore D}. This is followed by pausing the refresh operations for a certain amount of time (\emph{Refresh-pause interval}) and subsequent reading of the data (from the same block) containing the bit-flip patterns. This readout data, \emph{Bitstream\textunderscore R}, along with the identifier, \emph{Id}, comprises the (final) \comb response that is sent back to the authenticator [step \textcircled{\raisebox{-.9pt}7}]. Note that, in the present design, the operations in the \emph{SRAM Error Correction}, \emph{XOR}, and the \emph{HASH} stages are carried out in software (executing on the client processor) since the computational (and latency) overhead of these operations was observed to be very low. 

An interesting parameter introduced in \cite{dpuf_cases2} and also utilized in the present design is \emph{Wrapper pattern}. It specifies the peripheral data-bits that are written just before the beginning and after the end of the DRAM block, and influence the bit-flip patterns or responses from the DRAM. \emph{Wrapper pattern} can be one of several predefined types, \emph{e.g.,} all `1's, all `0's, checkered, \emph{etc}.

\begin{figure}
	\centering
 	\includegraphics[width=0.9\hsize]{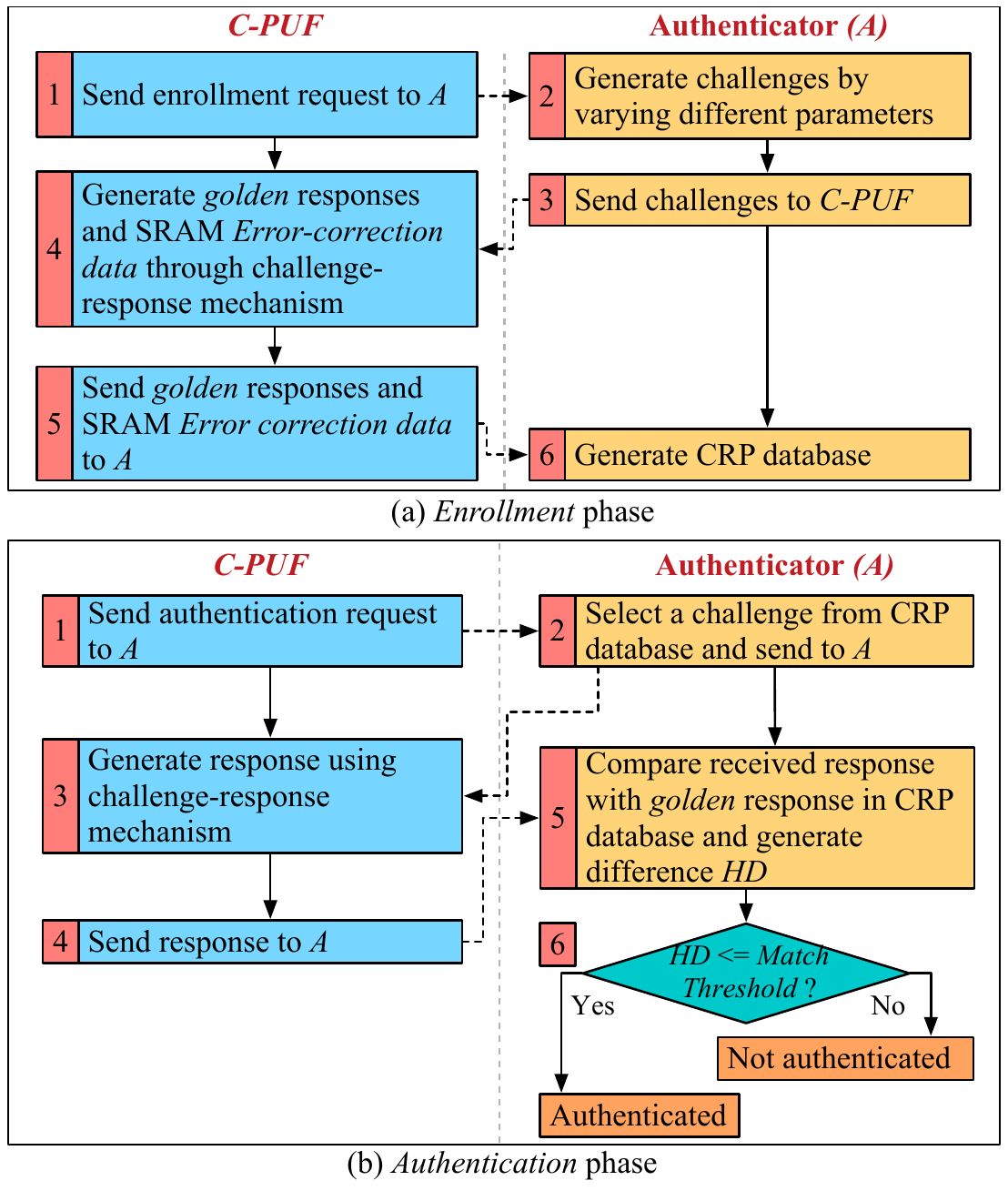}
	\vskip -5pt
	\caption{Different phases in authentication using \comb}
	\label{fig:flow}
	\vskip -17pt
\end{figure}

\vskip -15pt
\subsection{Authentication using C-PUF}
\label{cpuf_auth}
Two distinct phases are associated with an authentication process involving \comb~\textendash~\emph{enrollment} phase and \emph{authentication} phase, as shown in Fig.~\ref{fig:flow}. Both phases utilize the same challenge-response mechanism (described earlier) for response generation but differ in their objectives. The \emph{enrollment} phase primarily deals with the generation of the CRP database by subjecting \comb to different challenges and recording the generated responses. These responses serve as the \emph{golden} (expected) responses.
Also, during this phase, data for subsequent error correction is derived from the \emph{golden} start-up values of the SRAM (Algo.~\ref{algo:one}, described in Sec.~\ref{sec:err}).

Next, actual authentication of \comb happens during the \emph{authentication} phase, where it is subjected to a subset of the challenges (selected from the CRP database) and is expected to reproduce the \emph{golden} responses. Note that the error-correction data, generated during the \emph{enrollment} phase, is used here to correct the bit-errors in the SRAM start-up values (Algo.~\ref{algo:two}, described in Sec.~\ref{sec:err}). Finally, the proposed \comb design employs a fuzzy authentication strategy \cite{sram_puf_holcomb, puf_auth, dpuf_cases2} at the authenticator end to determine the outcome of the authentication process. At the core of this strategy is \emph{Match Threshold} (MT); \comb is successfully authenticated only if the Hamming Distance (HD) between the \emph{golden} response and the response generated by it during the \emph{authentication} phase is less than or equal to the MT value. This value is set by the authenticator based on the results obtained from the \emph{characterization} of \comb, which is described next.


\emph{C-PUF's} \emph{characterization} involves understanding its challenge-response behavior, which is affected by environmental factors (operating conditions) such as temperature, aging, \etc besides factors (parameters) specific to SRAM (\emph{Addr\textunderscore S} and \emph{Size\textunderscore S}) and DRAM (\emph{refresh-pause interval}, \emph{wrapper pattern}, \emph{etc}).
The \emph{characterization} process involves subjecting \comb to its challenge-response mechanism (described earlier) iteratively while varying the above-mentioned factors, and followed by an analysis of the generated responses. The vital insights gained from this analysis is used to make various design choices. For example, the MT value for \comb (under the specified operating conditions) is set as per the maximum value (in terms of HD) by which the responses deviate from the \emph{golden} responses. Specifically, for setting the MT value, we draw insights from~\cite{dpuf_cases2} that describes a similar methodology for DRAM PUFs.  Also, as in \cite{dpuf_cases2}, this analysis helps us in identifying blocks in DRAM that exhibit maximum entropy (bit-flips) at minimum \emph{refresh-pause intervals}.
Since the latency of response generation in \comb is primarily decided by \emph{refresh-pausing} in DRAM (SRAM \emph{power-cycling} is very fast), using a low \emph{refresh-pause interval} ensures an overall low operational latency. 

\begin{figure}[t]
	\centering
	\includegraphics[width=0.95\columnwidth]{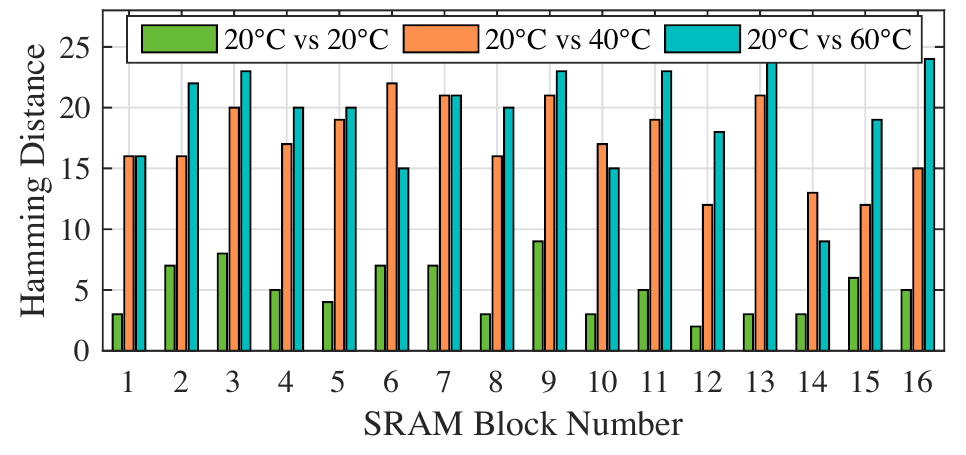}
	\caption{Variations in SRAM start-up values with temperature}
	\label{fig:sram_startup}
	\vskip -13pt
\end{figure}

\SetKwFunction{proc}{gen_err_corr_data}
\begin{algorithm}[t]
	\setlength{\textfloatsep}{0pt}
	\footnotesize
	\LinesNumbered
	\caption{\textbf{Generation of Error Correction Data}}
	\label{algo:one}
	\KwIn{$V\textsubscript{exp}$ = Golden (expected) start-up value from SRAM,\\
		$N$ = Number of bits in a segment 
	}
	\KwOut{$D$ = Error correction data\\
	}
	\BlankLine
	$D = \phi$\\
	$S = Get\_All\_NonOverlapSegments(V\textsubscript{exp})$\\
	\ForEach{$s \in S$}{
		$repBit\textsubscript{s} = 0$\\
		\If{$Num\_Ones(s) \geq Num\_Zeros(s)$}{
			$repBit\textsubscript{s} = 1$\\
		}
		$repSeg\textsubscript{s} = 0$\\
		\For{$i = 1$ \KwTo $N$}{
			$repSeg\textsubscript{s} = (repSeg\textsubscript{s} << 1) \mid repBit\textsubscript{s}$;\\
		}
		$D\textsubscript{s} = repSeg\textsubscript{s} \oplus s$\\
		$D = D \cup D\textsubscript{s}$
	} 
\end{algorithm}

\SetKwFunction{proc}{corr_startup_val}
\begin{algorithm}[t]
	\setlength{\textfloatsep}{0pt}
	\footnotesize
	\LinesNumbered
	\caption{\textbf{Error Correction and Generation of Corrected Start-up Value}}
	\label{algo:two}
	\KwIn{$V\textsubscript{err}$ = Erroneous start-up value from SRAM,\\
		$N$ = Number of bits in a segment,\\
		$D$ = Error correction data
	}
	\KwOut{$CV$ = Corrected start-up value\\
	}
	\BlankLine
	$CV = \phi$\\
	$S = Get\_All\_NonOverlapSegments(V\textsubscript{err})$\\
	\ForEach{$s \in S$}{
		$corSeg\textsubscript{s} = D\textsubscript{s} \oplus s$\\
		$repBit\textsubscript{s} = 0$\\
		\If{$Num\_Ones(corSeg\textsubscript{s}) \geq Num\_Zeros(corSeg\textsubscript{s})$}{
			$repBit\textsubscript{s} = 1$\\
		}
		$CV = CV \cup repBit\textsubscript{s}$ 
	}
\end{algorithm}

\subsection{Error Correction in SRAM}
\label{sec:err}
SRAM start-up values are affected by environmental and temporal variations, which could hinder \emph{C-PUF's} ability to perform authentication successfully. To demonstrate the impact of one such variation \emph{viz.} temperature, we generated start-up values from 16 different blocks (each 32 bytes in size) belonging to a SRAM and at three different temperatures - 20$^{\circ}$C, 40$^{\circ}$C, and 60$^{\circ}$C. Fig.~\ref{fig:sram_startup} shows the difference, in terms of HD, between the start-up values generated at the three temperatures for each of the blocks.
Since the proposed design utilizes a mathematical \emph{hash} (SHA-256) function to mask the SRAM start-up values, even a single bit-error could result in a completely different bitstream being subsequently applied to the DRAM (step \textcircled{\raisebox{-.9pt}6} in Fig.~\ref{fig:arch}). Therefore, the start-up values need to undergo perfect error-correction, \ie all bit-errors must be corrected before moving onto the next-stage (\emph{XOR}). We present two algorithms, Algo.~\ref{algo:one} and Algo.~\ref{algo:two}, that enable this in \comb while utilizing minimal computational and storage resources. 

Algo.~\ref{algo:one} is utilized during the \emph{enrollment} phase to generate the data that is subsequently used for correcting errors in SRAM start-up values. Note that the error correction data is always generated with respect to the \emph{golden} (expected) start-up value. Algo.~\ref{algo:one} starts by dividing the \emph{golden} start-up value into smaller \emph{segments}; a \emph{segment} comprises of a fixed number of bits (8-bits, used here). Each \emph{segment} is then assigned a \emph{representative bit-value} depending upon the relative number of \emph{one}-bits and \emph{zero}-bits in the \emph{segment}. The \emph{representative bit-value} is then expanded to form a \emph{representative segment}; the latter is \emph{xor}-ed with the \emph{segment} to generate the \emph{correction data} for that particular \emph{segment}. The \emph{correction data} from each \emph{segment} is then combined to form the final error correction data.

Algo.~\ref{algo:two} is utilized during the \emph{authentication} phase and uses the data generated earlier (during \emph{enrollment} phase) to perform error correction in SRAM start-up values. It starts by dividing the erroneous start-up value into smaller \emph{segments} of the same size (8-bits) as in Algo.~\ref{algo:one}. Each \emph{segment} is then \emph{xor}-ed with its respective \emph{correction data} to generate the \emph{corrected segment}. The relative number of \emph{one}-bits and \emph{zero}-bits in a \emph{corrected segment} decides its \emph{representative bit-value}. All the {representative bit-values} are then combined to form the corrected start-up value (\emph{CV}).


\vskip -15pt
\subsection{Intrinsic Reconfigurability}
\label{sec:des_reconf}
Reconfigurability refers to the ability of a PUF to undergo reconfiguration, \ie modify its challenge-response behavior~\cite{reconf_puf}. The new behavior is unpredictable and cannot be modeled based on the knowledge of the behavior prior to reconfiguration, giving the PUF substantial protection against various attacks \cite{ml_attack}. The proposed design achieves reconfigurability in \comb \emph{intrinsically}, \ie without using any additional resource, unlike~\cite{reconf_puf}. It specifies two knobs for reconfiguration \textendash~\emph{Addr\textunderscore S} and \emph{Refresh-pause interval}. Changing \emph{Addr\textunderscore S} in the challenge-response mechanism generates (new) start-up values from a different block in the SRAM. On the other hand, modifying the \emph{Refresh-pause interval} generates new bit-flip patterns from the (same) DRAM block. Hence, by turning one or both reconfiguration knobs, \comb can undergo reconfiguration \emph{intrinsically} and start behaving as a new PUF. Note that, unlike other parameters in the challenge (Fig.~\ref{fig:arch}), \emph{Addr\textunderscore S} and \emph{Refresh-pause interval} are reserved solely for reconfiguration, and hence do not vary across different challenges during authentication runs, except when there is a need for reconfiguration.
 \section{Experimental Setup}
\label{sec:exp}
\begin{figure}[t]
	\centering
	\includegraphics[width=1\columnwidth]{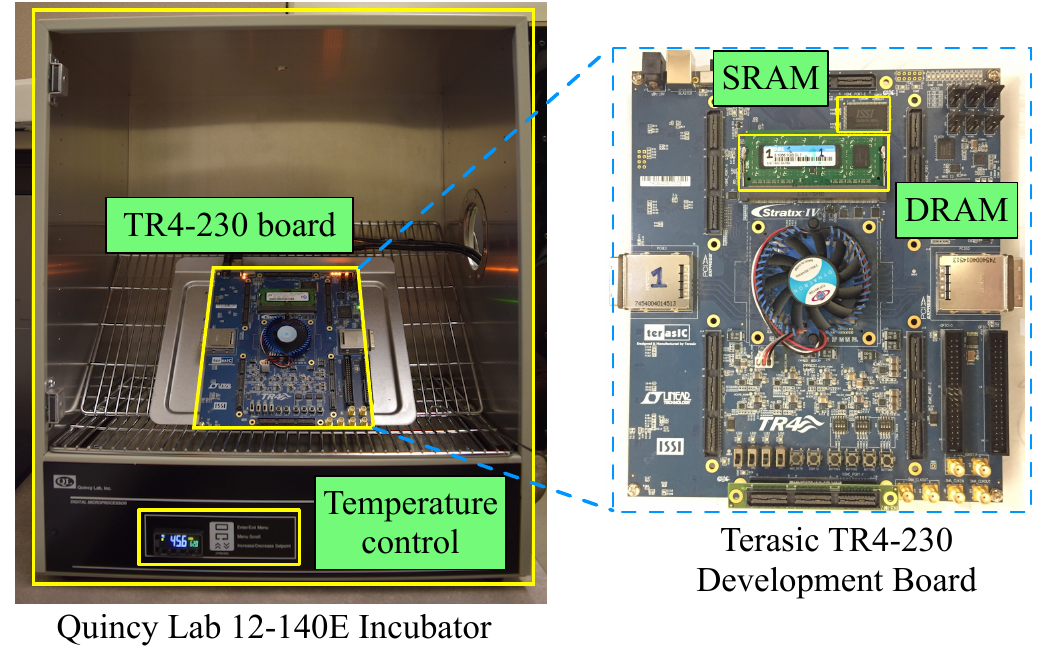}
	\vskip -5pt
	\caption{Photograph of our experimental setup}
	\label{fig:exp}
	\vskip -15pt
\end{figure}

This section provides a brief description of the experimental setup used to validate the \comb design. It consists of a Terasic TR4-230 development board \cite{tr4}, containing an Altera Stratix IV GX FPGA, 2MB SSRAM (Synchronous SRAM), and 1GB DDR3 DRAM (SODIMM). The temperature and aging experiments were performed by operating the TR4-230 development board inside the Quincy Lab 12-140E Incubator. Fig.~\ref{fig:exp} shows the complete experimental setup. 

The FPGA was programmed with a soft Nios II processor~\cite{nios} along with an Altera Generic Tri-State Controller and an Altera UniPHY DDR3 memory controller for controlling the SRAM and DRAM modules, respectively. A custom slave running on the processor was also created, which can instruct the memory controller to pause the DRAM refresh operations. To keep the design simple, the SRAM start-up values were generated by \emph{power-cycling} the whole development board and subsequently reading the contents of the SRAM. 
A total of five \comb instances were constructed for validation, each consisting of a SSRAM and DDR3 DRAM SODIMM. While the SSRAMs belonged to two different manufacturers, the DRAMs were procured from five different manufacturers.
%

 \section{Results}
\label {sec:res}

\begin{figure}[t]
	\centering
	\includegraphics[width=1\columnwidth]{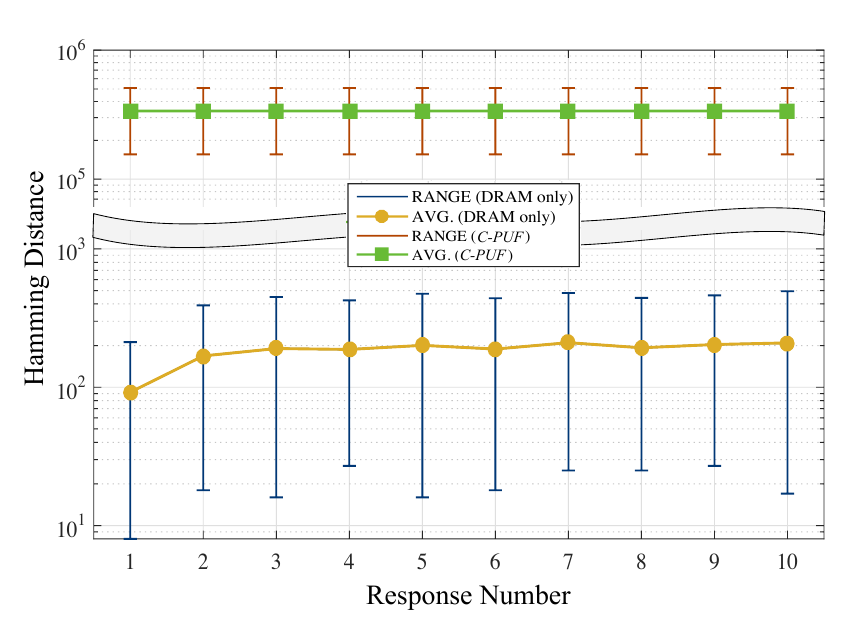}
	\vskip -5pt
	\caption{Enhanced uniqueness obtained using \comb compared to DRAM PUFs}
	\label{fig:uniq}
	\vskip -13pt
\end{figure}

\begin{figure}[t]
	\centering
	\includegraphics[width=\columnwidth]{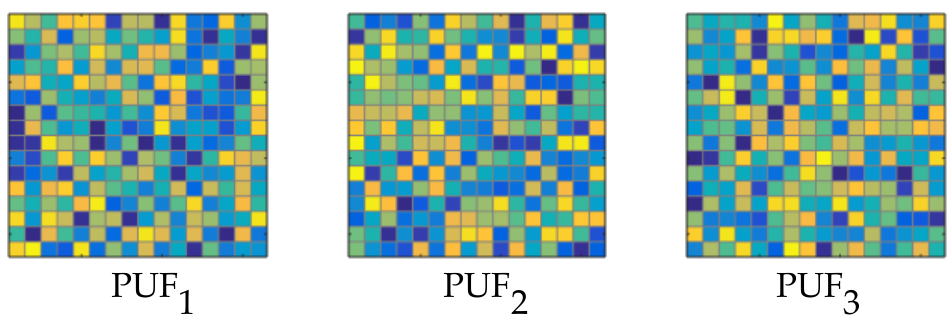}
	\vskip -5pt
	\caption{Responses from different \cinss}
	\label{fig:fingerprint}
	\vskip -15pt
\end{figure}

\begin{table*}[t]
	\centering
	\small
	\caption{Parameter values used in our experiments}
	\begin{tabular}{|c|c|c|c|c|c|c|}
		\hline
		\bf{Addr\textunderscore S}&\bf{Size\textunderscore S}&\bf{Addr\textunderscore D}&\bf{Size\textunderscore D}&\bf{Bitstream\textunderscore C}&\bf{Wrapper pat.}&\bf{Refresh-pause int.}\\
		\hline	
		Varied&32 B&Varied&128 KB&Varied&All `1's&40 sec\\
		\hline
	\end{tabular} 
	\label{param_table}
	\vskip -10pt
\end{table*}
This section presents the results obtained from experiments conducted to validate our work.
Table~\ref{param_table} provides a summary of the parameter values (in challenges) used in the experiments.
\vspace{-0.01in}
\subsection{Uniqueness Analysis}
\label{uniq_any}
The ability of a PUF in generating unique responses forms the very foundation of challenge-response-based authentication.
Hence, to demonstrate the uniqueness of the responses generated by the proposed design, five \cinss (PUF\textsubscript{1}\textendash PUF\textsubscript{5}) were each subjected to ten different (and random) challenges at 20$^{\circ}$C. For every challenge, the responses generated by the \inss were compared against each other, and the differences are plotted as HD in Fig.~\ref{fig:uniq} (top). As described in Sec.\ref{sec:des}, \comb generates the responses by \emph{xor}-ing (in \emph{XOR} stage) and \emph{hash}-ing (in \emph{HASH} stage) the corrected start-up value of the SRAM and subsequently applying it to the DRAM. Hence, one may argue that the uniqueness of the responses generated by \comb is contributed by the \emph{hash} (SHA-256) operation only and not by the rest of the design. However, this is not the case, as shown in Fig.~\ref{fig:uniq}, where the responses were generated by skipping the \emph{hash} operation (\emph{HASH} stage) altogether. In other words, the responses of the \cinss were generated by directly applying the \emph{xor}-ed start-up values of the SRAM to the DRAM. As shown, the minimum HD across all ten responses is greater than $155,000$, and hence the responses are truly unique. Fig.~\ref{fig:fingerprint} gives a pictorial representation of the responses generated by three of the \cinss when subjected to the same challenge. 
Note that applying the \emph{hash} operation (during a different experiment), in fact, further enhanced the uniqueness of the generated responses by bringing the minimum HD very close to the ideal value ($524,288$). 

As mentioned earlier, an advantage with \comb is the enhanced uniqueness provided by it as compared to a DRAM-only PUF \cite{dpuf_cases2,dram_puf_keller,commodity_dpuf}. To demonstrate this, five DRAM-only PUFs were constructed (as per the design in \cite{dpuf_cases2}) using the DRAMs extracted from the \cinss. Next, each was subjected to the same ten challenges that were previously applied to the \cinss.
As shown in Fig.~\ref{fig:uniq} (bottom), comparison of the responses generated by these DRAM-only PUFs yielded a maximum HD of $494$, which is three orders of magnitude less than that of the \cinss.
 
%
\subsection{Robustness Analysis: Authentication under Temperature Variations and Aging Effects}
\label{res:robust}
Robustness refers to a PUF's ability to undergo successful authentication under different operating conditions, primarily determined by \emph{temperature} and \emph{aging}. We present a robustness analysis of the proposed \comb design below.

\subsubsection{Authentication under Temperature Variations}
\begin{figure}
	\centering
	\includegraphics[width=1\columnwidth]{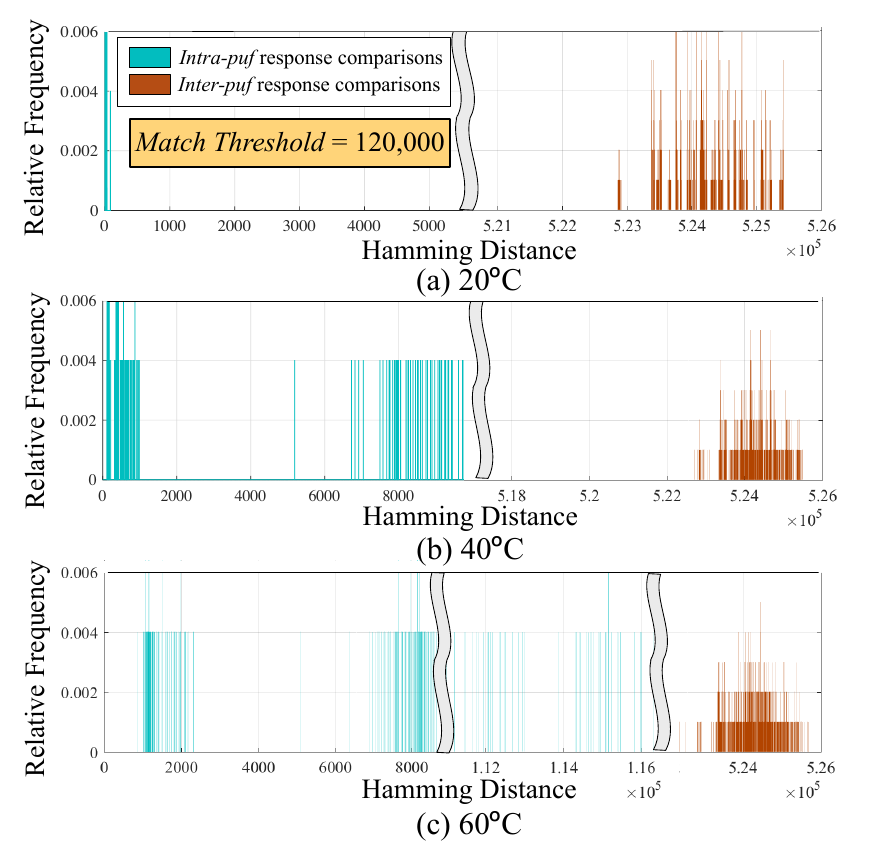}
	\vskip -5pt
	\caption{Authentication under temperature variations}
	\label{fig:auth_temp}
	\vskip -13pt
\end{figure}

To demonstrate the robustness of the design to temperature variations, the \cinss were made to undergo \emph{enrollment}  (Fig.~\ref{fig:flow}(a)) at 20$^{\circ}$C, which involved applying fifty different (and random) challenges to each of the \cinss. The generated responses served as the \emph{golden} responses, and were stored in the CRP database. Next, to emulate an actual scenario, the authentication (Fig.~\ref{fig:flow}(b)) was performed at three different temperatures \textendash~20$^{\circ}$C, 40$^{\circ}$C, and 60$^{\circ}$C, by reapplying the same challenges to the \inss.
At each temperature, the responses of the \cinss generated during authentication were compared with their respective \emph{golden} responses to calculate the \emph{intra-puf} comparison HD, as shown in Fig.~\ref{fig:auth_temp}.  Note that the \emph{y-axis} represents \emph{relative frequency}, \ie the fraction of the total comparisons, either \emph{intra-puf} or \emph{inter-puf} (explained below), that yields a certain HD (\emph{x-axis}). As evident, setting the \emph{match threshold} value to $120,000$ successfully authenticated all the \cinss with respect to every challenge and operating temperature, thus achieving a $100\%$ true-positive rate.

We also performed \emph{inter-puf} comparisons, \ie the golden responses of a \cins were compared with the responses generated during authentication of every other instance. Fig.~\ref{fig:auth_temp} shows the \emph{relative frequency} versus HD for the \emph{inter-puf} comparisons corresponding to the same fifty challenges and operating temperatures. At each temperature, the wide HD margin between the \emph{intra-puf} and \emph{inter-puf} comparisons ensured the absence of any false-positives, thereby achieving a $0\%$ false-positive rate. 

\begin{figure}
	\centering
	\includegraphics[width=0.85\columnwidth]{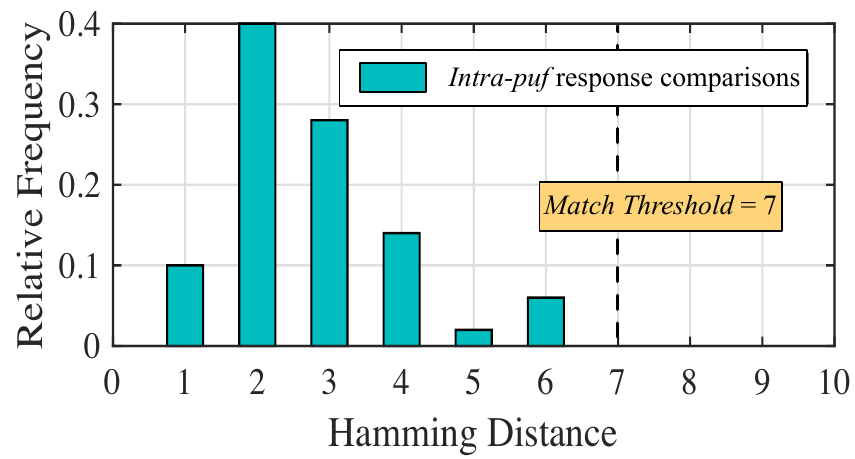}
	\vskip -5pt
	\caption{Authentication under aging effects}
	\label{fig:auth_aging}
	\vskip -13pt
\end{figure}

\subsubsection{Authentication under Aging Effects}
To demonstrate \emph{C-PUF's} robustness to temporal variations or aging affects, \emph{enrollment} of one of the \cinss was carried out at 20$^{\circ}$C by applying ten different (and random) challenges to it and subsequently recording the corresponding (\emph{golden}) responses. Next, the instance was subjected to an accelerated aging process by applying a temperature of 85$^{\circ}$C for 48 hours that effectively aged it by 12 months \cite{aging_test}. Authentication was then performed by generating the responses (with the same challenges and at the same temperature) from the aged instance; the responses were then compared with the golden ones (\emph{intra-puf} comparisons). Fig.~\ref{fig:auth_aging} shows the \emph{relative frequency} versus HD for these comparisons. As evident, setting the \emph{match threshold} value to $7$ successfully authenticated the \ins for every challenge, thus achieving a $100\%$ true-positive rate without any false-positives. 
Note that the \emph{inter-puf} comparisons are not depicted in Fig.~\ref{fig:auth_aging} as the corresponding HD values are very high ($>500,000$).
 \section{Discussions}
\label{sec:dis}
We now provide some additional design aspects of \comb.
\subsection{Security Analysis}
We envision two types of attacks on \comb \textendash~invasive \cite{invasive} and non-invasive \cite{ml_attack}. The utilization of heterogeneous memory technologies (entropy sources), which are spatially distributed (on-chip and off-chip), substantially improves \emph{C-PUF's} resistance to invasive attacks. On the other hand, the presence of multiple variable parameters in \emph{C-PUF's} challenge-response mechanism generates an exponential number of CRPs. This, coupled with its ability to undergo reconfiguration, makes non-invasive attacks very difficult to mount. 

An alternative design (\emph{SD}) could have both an SRAM PUF and a DRAM PUF present in a device but operating independently of each other. \comb scores over such a design by supporting a comparatively much larger number of CRPs, thereby providing stronger defense against non-invasive attacks. Assuming a typical SRAM size (2 MB) and parameter values as specified in Table~\ref{param_table}, the SRAM PUF (alone) supports $A$ CRPs, where $A=2^{16}$. 
Similarly, assuming a typical DRAM size (1 GB) and three \emph{Wrapper patterns} and \emph{Refresh-pause intervals} each to choose from (as in \cite{dpuf_cases2}) as well as a fixed \emph{Bitstream\textunderscore C} (to keep this analysis simple), the DRAM PUF (alone) supports $B$ CRPs, where $B=9\times2^{13}$ or $\sim2^{16}$. 
As a result, while \emph{SD} supports a total of $A+B$ ($2\textsuperscript{17}$) CRPs, the number of CRPs supported by \comb is of the order of $A\times B$ ($2\textsuperscript{32}$), which is multiple orders of magnitude higher. In reality, \emph{Bitstream\textunderscore C} is also varied alongside \emph{Size\textunderscore S} and \emph{Size\textunderscore D} (for generating random challenges), thus increasing this difference in the number of CRPs even further.
   

\subsection{Replacing Genuine DRAM with a Counterfeit: A Case-Study}
\label{dram_err}

Several embedded systems contain DRAMs in the form of Dual In-line Memory Modules (DIMMs) that (unlike SRAMs, which are physically soldered) could be easily detached from the system. Hence, an attacker with physical access to the system may be able to mount an invasive attack by replacing the genuine DIMMs with counterfeit ones. Ideally, the now compromised system should not be authenticated, however, the version of \comb described here may authenticate it as a genuine system. This is because, to highlight the core competencies of \comb within the paper's limited space, the support mechanism, which includes setting the appropriate \emph{match threshold} (MT) value, was kept relatively simple. In the described version, the MT value that decides the authentication outcome is set \emph{statically} (with the same value) \cite{dpuf_journal} for multiple \cinss, and is also independent of the operating conditions (\eg temperature). Hence, if the HD from the response comparisons is within the MT, the system may undergo successful authentication in spite of containing a counterfeit DIMM. To protect against such a scenario (attack) an enhanced version of \comb was also developed that sets the MT value through the framework specified in \cite{dpuf_journal}, which takes several factors such as behavior of the particular DRAM, extent of variations in the operating conditions, \etc into consideration to come up with a \emph{dynamic} DRAM-specific MT. Note that the enhanced \comb version prevented the system with the counterfeit DIMM from getting authenticated.

 \section{Related Work}
\label{sec:relwork}
A variety of memory-based PUFs has been proposed over the years. Ref.~\cite{sram_puf_holcomb} extracted unique fingerprints and generated true random numbers by using the start-up values of SRAM. Device authentication using DRAM refresh pausing under wide environmental and temporal variations was presented in~\cite{dpuf_cases2}. Another approach to PUF design targeted achieving reconfigurability in PUFs~\cite{reconf_puf}. Ref.~\cite{flash_puf_prabhu} proposed mechanisms to extract device fingerprints from {\sc Flash}.

The idea of using multiple memories to derive unique keys from an electronic system was first mentioned in \cite{dist_puf}. 
However, to the best of our knowledge, no physical implementation or results thereof have been published yet. 
A similar but non-memory-based work was presented in \cite{super_puf}, which proposes to combine on-chip entropy sources (\eg clock sinks) to generate CRPs. 
Unlike these works, \comb does not require the addition of custom hardware to the existing circuitry, and hence can be implemented easily using COTS devices.         
 \section{Conclusion}
We proposed the concept and design of a memory-based combination PUF that intelligently combines two memory technologies, SRAM and DRAM, to overcome several shortcomings of current memory-based PUFs. Extensive tests conducted on a real implementation of the PUF demonstrate the robustness of the proposed design, achieving a 100\% true-positive rate and 0\% false-positive rate during authentication.  In future, we plan to include more device components in the \comb design as well as analyze its performance under supply-voltage variations.





%

\bibliographystyle{unsrt}
{
\bibliography{combpuf_bib_short}  

\begin{thebibliography}{10}

\bibitem{invasive}
C.~Helfmeier et~al.
\newblock {Cloning Physically Unclonable Functions}.
\newblock In {\em HOST}, 2013.

\bibitem{ml_attack}
U.~R\"{u}hrmair et~al.
\newblock Modeling attacks on physical unclonable functions.
\newblock In {\em CCS}, 2010.

\bibitem{bogus}
M.~Pecht et~al.
\newblock {Bogus: electronic manufacturing and consumers confront a rising tide
  of counterfeit electronics}.
\newblock {\em IEEE Spectrum}, May 2006.

\bibitem{dev_auth}
G.~E. Suh et~al.
\newblock {Physical Unclonable Functions for Device Authentication and Secret
  Key Generation}.
\newblock In {\em DAC}, 2007.

\bibitem{dpuf_cases2}
S.~Sutar et~al.
\newblock {D-PUF: An intrinsically reconfigurable DRAM PUF for device
  authentication in embedded systems}.
\newblock In {\em CASES}, 2016.

\bibitem{sram_puf_holcomb}
D.~E. Holcomb et~al.
\newblock {Power-Up SRAM State as an Identifying Fingerprint and Source of True
  Random Numbers}.
\newblock {\em IEEE Transactions on Computers}, Sept 2009.

\bibitem{dram_puf_keller}
C.~Keller et~al.
\newblock {Dynamic memory-based physically unclonable function for the
  generation of unique identifiers and true random numbers}.
\newblock In {\em ISCAS}, 2014.

\bibitem{commodity_dpuf}
W.~Xiong et~al.
\newblock {Run-time Accessible DRAM PUFs in Commodity Devices}.
\newblock In {\em CHES}, 2016.

\bibitem{flash_puf_prabhu}
P.~Prabhu et~al.
\newblock {Extracting device fingerprints from flash memory by exploiting
  physical variations}.
\newblock In {\em TRUST}, 2011.

\bibitem{dist_puf}
P.~T. Tuyls et~al.
\newblock {Distributed PUF}, 2014.
\newblock {US Patent 8,699,714}.

\bibitem{super_puf}
M.~Wang et~al.
\newblock {SuperPUF: Integrating heterogeneous Physically Unclonable
  Functions}.
\newblock In {\em ICCAD}, 2014.

\bibitem{mc_spuf}
C.~B\"{o}hm et~al.
\newblock {A microcontroller SRAM-PUF}.
\newblock In {\em NSS}, 2011.

\bibitem{puf_auth}
W.~Che et~al.
\newblock {PUF-Based Authentication}.
\newblock In {\em ICCAD}, 2015.

\bibitem{reconf_puf}
I.~Eichhorn et~al.
\newblock {Logically Reconfigurable PUFs: Memory-based Secure Key Storage}.
\newblock In {\em STC}, 2011.

\bibitem{tr4}
Terasic.
\newblock {TR4 FPGA Development Kit}, March 2015.

\bibitem{nios}
Altera.
\newblock {Nios II} processor, March 2015.

\bibitem{aging_test}
A.~Maiti et~al.
\newblock {The Impact of Aging on a Physical Unclonable Function}.
\newblock {\em IEEE Transactions on VLSI Systems}, Sept 2014.

\bibitem{dpuf_journal}
{DRAM PUF for Device Authentication and Random Number Generation}.
\newblock In {\em ACM Transactions on Embedded Computing Systems}, Accepted for
  publication, Nov 2017.

\end{thebibliography}
}
\end{document}